\documentclass[twocolumn,switch]{article} 
\usepackage{preprint}

\usepackage[numbers,square]{natbib}
\usepackage[utf8]{inputenc}	
\usepackage[T1]{fontenc}	
\usepackage{xcolor}		
\usepackage[colorlinks = true,
            linkcolor = purple,
            urlcolor  = blue,
            citecolor = cyan,
            anchorcolor = black]{hyperref}	
\usepackage{booktabs} 		
\usepackage{nicefrac}		
\usepackage{microtype}		
\usepackage{lineno}		
\usepackage{float}			
\usepackage{multicol}		

\usepackage{lipsum}		
\usepackage{orcidlink}

\usepackage{newfloat}
\DeclareFloatingEnvironment[name={Supplementary Figure}]{suppfigure}
\usepackage{sidecap}
\usepackage{longtable}
\usepackage{array}
\usepackage{tabularx}
\sidecaptionvpos{figure}{c}
\usepackage{titlesec}
\titlespacing\section{0pt}{12pt plus 3pt minus 3pt}{1pt plus 1pt minus 1pt}
\titlespacing\subsection{0pt}{10pt plus 3pt minus 3pt}{1pt plus 1pt minus 1pt}
\titlespacing\subsubsection{0pt}{8pt plus 3pt minus 3pt}{1pt plus 1pt minus 1pt}

\usepackage{circledsteps}
\usepackage{tikz,xcolor,hyperref}
\usepackage{balance}
\newcolumntype{R}[1]{>{\raggedleft\let\newline\\\arraybackslash}p{#1}} 
\newcolumntype{L}[1]{>{\raggedright\let\newline\\\arraybackslash}p{#1}} 
\newcolumntype{C}[1]{>{\centering\let\newline\\\arraybackslash}p{#1}} 

\newcommand{\citesec}[1]{Section~\ref{sec:#1}}

\title{ FORTE: An Open-Source System for Cost-Effective and Scalable Environmental Monitoring}

\usepackage{eso-pic} 

\AddToShipoutPictureBG*{%
  \ifnum\value{page}=1
    \put(0,20){%
      \makebox[\paperwidth]{\hfill \small\tt{*correspondence: Zoe.Pfister@uibk.ac.at} \hfill}%
    }%
  \fi
}
\usepackage{authblk}

\author[1\thanks{\tt{Zoe.Pfister@uibk.ac.at}}]{Zoe Pfister \orcidlink{0009-0009-2882-5059}}
\author[1]{Michael Vierhauser \orcidlink{0000-0003-2672-9230}}
\author[2]{Alzbeta Medvedova \orcidlink{0009-0001-3792-8908}}%
\author[2]{Marie Schroeder \orcidlink{0000-0002-4421-6482}}%
\author[2]{Markus Rampp}
\author[5]{Adrian Kronenberg}
\author[3]{Albin Hammerle \orcidlink{0000-0003-1963-5906}}%
\author[3]{Georg Wohlfahrt \orcidlink{0000-0003-3080-6702}}%
\author[1]{Alexandra Jäger}
\author[1]{Ruth Breu}
\author[4]{Alois Simon \orcidlink{0000-0002-6718-7354}}%

\affil[1]{University of Innsbruck\\Department of Computer Science\\Technikerstra\ss e 21a\\6020 Innsbruck\\Austria}
\affil[2]{University of Innsbruck\\Department of Atmospheric and Cryospheric Sciences\\Innrain 52f\\6020 Innsbruck\\Austria}
\affil[3]{University of Innsbruck\\Department of Ecology\\Sternwartestra\ss e 15\\6020 Innsbruck\\Austria}
\affil[4]{Amt der Tiroler Landesregierung, Gruppe Forst\\Abt. Forstplanung\\Bürgerstra\ss e 36\\6020 Innsbruck\\Austria}
\affil[5]{TU Wien\\Faculty of Physics\\Wiedner Hauptstra\ss e 8-10\\1040 Wien\\Austria}

\begin{document}

\twocolumn[ 
  \begin{@twocolumnfalse} 
  
\maketitle

\begin{abstract}
Forests are an essential part of our biosphere, regulating climate, acting as a sink for greenhouse gases, and providing numerous other ecosystem services. However, they are negatively impacted by climatic stressors such as drought or heat waves. In this paper, we introduce FORTE, an open-source system for environmental monitoring with the aim of understanding how forests react to such stressors. It consists of two key components: (1) a wireless sensor network (WSN) deployed in the forest for data collection, and (2) a Data Infrastructure for data processing, storage, and visualization. The WSN contains a Central Unit capable of transmitting data to the Data Infrastructure via LTE-M and several spatially independent Satellites that collect data over large areas and transmit them wirelessly to the Central Unit. Our prototype deployments show that our solution is cost-effective compared to commercial solutions, energy-efficient with sensor nodes lasting for several months on a single charge, and reliable in terms of data quality. FORTE's flexible architecture makes it suitable for a wide range of environmental monitoring applications beyond forest monitoring. The contributions of this paper are three-fold. First, we describe the high-level requirements necessary for developing an environmental monitoring system. Second, we present an architecture and prototype implementation of the requirements by introducing our FORTE platform and demonstrating its effectiveness through multiple field tests. Lastly, we provide source code, documentation, and hardware design artifacts as part of our open-source repository.
\end{abstract}
\vspace{0.35cm}

  \end{@twocolumnfalse} 
] 

\section{Introduction} 
\label{sec:introduction}

Forests provide several important key ecosystem services, such as harboring large biodiversity, regulating climate, acting as a sink for greenhouse gases, and providing timber resources~\cite{royIncreasingForestLoss2014}. At the same time, forests are negatively impacted by various stressors such as climate change and pests. To better understand how forests respond to these stressors, various data of forest ecosystems needs to be collected and analyzed, typically over extended periods of time~\cite{lauschUnderstandingForestHealth2017,porterWirelessSensorNetworks2005}. This includes, for example, data on forest interior climate, soil moisture, and tree growth response. While several commercial solutions for environmental monitoring are available, these are typically expensive, proprietary, and only allow for limited spatial coverage. 

To address these shortcomings of existing solutions, together with the Forest Service of the Office of the Tyrolean Government, we have developed a novel open-source forest measurement and analysis platform called FORTE. The key innovation of FORTE is the hardware design of the measurement stations, consisting of a Central Unit (CU) capable of transmitting data to a remote server and several spatially separated, modular, low-power Satellites that collect data over large areas and transmit them wirelessly to the CU. In addition, our platform includes a remote Backend for data storage, quality analysis, and visualization. The platform is a result of a multi-year interdisciplinary collaboration between experts from computer science, ecology, and forest management. During the development of FORTE, the primary objective was to help forest managers, researchers, and forest owners to monitor the health and productivity of their forests over time with the goal to being alerted and responding to impending disturbances, and allow impact assessments of environmental events. FORTE was designed to enable cost-effective deployment of multiple measurement stations and data loggers across forest locations. However, the exact choice of sensors is flexible, allowing the system to be easily adjustable for other, similar use cases, such as climate or air quality monitoring.

The contributions of this paper are three-fold. After describing related work (\citesec{sota}), we first specify the core requirements necessary for developing an environmental monitoring system (\citesec{materials-and-methods}). Second, based on these requirements, we present a high-level architecture and prototype implementation of our FORTE system (\citesec{forte-infrastructure}), and third, we demonstrate its effectiveness through field tests in forest environments (\citesec{evaluation}). 
Additionally, the source code, documentation, and hardware-design files are provided as part of our open-source repository located at \url{https://git.uibk.ac.at/informatik/qe/forte}.
\section{Related Work}
\label{sec:sota} 

The concept of developing wireless sensor networks (WSNs)~\cite{matinOverviewWirelessSensor2012} for the purpose of monitoring the environment has been around for over two decades, with early works suggesting progress even before 2005~\cite{porterWirelessSensorNetworks2005}. Advances in technology, coupled with the increasing availability and affordability of Internet of Things (IoT) devices, have led to multiple research projects~\cite{corkeEnvironmentalWirelessSensor2010,liuDoesWirelessSensor2013,wageleMultisensorStationAutomated2022} and commercial products.

In 2010, Corke et al.~\cite{corkeEnvironmentalWirelessSensor2010} described various environmental monitoring applications for WSNs, including rainforest monitoring~\cite{warkSpringbrookChallengesDeveloping2008}. In contrast to FORTE, their solution was capable of multi-hop routing of data, which can result in a network with larger coverage but requires more energy. In FORTE, we decided against using a multi-hop WSN to make our sensor nodes as energy efficient as possible, as they may be placed under the canopy of trees where solar power does not provide sufficient energy for sustained operation. Another multi-hop WSN specifically for forest environment monitoring is GreenOrbs~\cite{liuDoesWirelessSensor2013}. GreenOrbs stands out in literature because of its large size of 330 sensor nodes, offering a large area of monitoring. The study found that mesh-routing using hundreds of sensor-nodes results in severe packet losses and overall network degradation. In a recent perspective paper, Wägele et al.~\cite{wageleMultisensorStationAutomated2022} proposed a network of Automated Multisensor stations for Monitoring of species Diversity, consisting of sensor nodes connected to base stations for power supply and data transmission. The base station transmits collected data to a Backend via an existing mobile network or satellite link. Their system is similar to our FORTE system, however, it is conceptualized specifically for biodiversity monitoring using specialized sensors whereas FORTE is suitable for general environmental monitoring.

Besides environmental monitoring, WSNs are used for other purposes, such as disaster monitoring or agriculture monitoring. Dampage et al.~\cite{dampageForestFireDetection2022} developed a Wireless Sensor Network system to detect forest fires using temperature, humidity, light intensity level, and CO level data, in combination with machine learning algorithms. In contrast to FORTE, disaster monitoring comes with different use cases, particularly in terms of data collection intervals. M.S. Koushik et al.~\cite{sDesignDevelopmentWireless2021a} developed a prototype WSN to collect agricultural data from fields in India, collecting air humidity, air temperature, and soil moisture data from sensor nodes wirelessly connected to a base station, referred to as sink. 

\section{Materials and Methods} 
\label{sec:materials-and-methods} 

Building a system for environmental monitoring requires a combination of
multiple subsystems, each of which has its own set of functional and
non-functional requirements. During initial stakeholder meetings, we
defined four primary groups of actors that will interact with the FORTE
system: (i) Forest Owners, who must be contacted for deployment
permission and informed about the system and its characteristics; (ii)
Foresters, who also need to be notified about upcoming and existing WSN
deployments in their forest area; (iii) Administrators, responsible for
setup and maintenance, who are also capable of viewing audit logs and
managing the system's users and adding or removing devices and sensors
to the FORTE system; and finally (iv) ``Users'', i.e., persons such as
researchers and forest managers, capable of viewing and downloading the
data collected by a deployment of the FORTE system.

Additionally, we specified the core components of the system, namely:
\emph{An in-forest, environmental measurement station (FORTE-WSN)}
separated into two parts: (1) a Central Unit (CU), providing long-range
communication capabilities to an external server, and (2) several
spatially independent Satellites that take individual measurements at
points of interest, and a \emph{Data Infrastructure} with a server-side
Backend for data validation and storage, and a Frontend for system
configuration, data viewing, and data retrieving.

\subsection{Method}\label{method}

During the development of FORTE, we followed an iterative, action
research-based approach, described as ``a process of systematic inquiry,
usually cyclical, conducted by those inside a community {[}\ldots{]};
its goal is to identify action that will generate improvement the
researchers believe important''~\cite{ivankova2015mixed}. A cycle in an action
research project is structured as follows: (1) a diagnosing step to
identify a problem, (2) an action planning step, (3) an action taking
step, (4) an evaluation step, where data about the action is collected
and analyzed, and finally (5) a learning step where new knowledge is
consolidated and documented. Learnings from one cycle can inform and
impact the decisions that will be made in the next iteration cycle~\cite{staron2020Action}.

The development of the FORTE system was split into two approximately
12-month-long cycles. The first cycle started with an initial
stakeholder workshop, where we established the goals of the project.
During the action planning step, we defined the initial set of
requirements, including which hardware to use for our sensor nodes.
Next, we built initial prototypes of our CU and Satellites, as well as
the first version of a server for end-to-end testing. We evaluated this
prototype through a set of lab experiments and a short-term field test.
In this first cycle, we made the three core observations that we
required (1) more precise sensors, (2) more reliable sensor connections
at the Satellites, and (3) better power management.

In the second cycle, we replaced several sensors, revisited the
electronic circuits and designed more reliable loggers using custom-made
PCBs to improve sensor connections and energy consumption. Further, we
defined data validity and UI requirements of the Data Infrastructure.
Based on this, we implemented the server and Frontend of the Data
Infrastructure, solved the sensor-connectivity issues, and improved the
wireless connectivity of both the CU and Satellites to reduce energy
consumption. This resulted in the creation of a second prototype (cf.
\citesec{demonstrator-setup}), three of which were evaluated in the field over a span of
three months. During the evaluation, we identified and subsequently
resolved an issue on the Satellite PCBs that resulted in the collection
of inaccurate data. After the field-test evaluation concluded in late
2023, we documented our learnings in a final report, which is available
in German at \url{https://dafne.at/projekte/forte}.

\subsection{Requirements}\label{requirements}

We separated the FORTE system into a Data Infrastructure and a WSN that
operates in the field (i.e., within a forest). In this section, we
present the key requirements (cf. Table~\ref{tab:1-key-requirements-cu-satellites} and Table~\ref{tab:2-key-requirements-data-infrastructure}) for the system~\cite{pfisterConceptsImplementationWireless2024}. Starting with the requirements defined for the FORTE-WSN
system listed in Table~\ref{tab:1-key-requirements-cu-satellites}, we require both the devices on the CU and the
Satellites to be highly energy efficient. More specifically, the
Satellite nodes, which are spatially separated from the CU, need to have
runtimes exceeding months, without the need to replace batteries. For
the CU, we relax this runtime duration to a week, since it continuously
listens for new data from Satellites and includes mobile networking
equipment for LTE-M data transmission, thus requiring significantly more
power. However, additional energy harvesting solutions such as solar
panels shall be installed at the CU to increase uptime. Since FORTE-WSN
operates in outdoor environments, e.g., in a forest, it necessitates
appropriate protection of its electronic components. Another requirement
was to keep the cost of materials besides sensors to a minimum, to allow
easy and widespread adoption. Naturally, both the CU and the Satellites
must be equipped with wireless transmission capabilities to communicate
both within the network in the field and, in the case of the CU, the
Backend server. We specify that the transmission technology in the field
must guarantee a range of at least 50 meters. To improve data
reliability, we back up measurement data during connection issues.
Lastly, the sensor nodes responsible for data collection in the
Satellites and the CU must be capable of interacting with a variety of
both digital and analog sensors. This includes interoperability with
sensors using wired communication technologies like
I\textsuperscript{2}C, RS-485, SDI-12, 1-wire, or analog.

\begin{table}[]
\centering
\caption{Key Requirements for the Central Unit and Satellites.}
\label{tab:1-key-requirements-cu-satellites}
\begin{tabularx}{\columnwidth}{lX}
\toprule
\textbf{Nr.} & \textbf{Central Unit (CU) and Satellites}                                                 \\ \midrule
1   & Both 			the CU and the Satellites shall be highly energy efficient.                 \\
2   & The 			enclosures of the CU and the Satellites shall be IP65 			weatherproof.       \\
3   & The 			hardware excluding the sensor equipment shall be affordable.                 \\
4   & The 			CU and Satellites shall have wireless transmission capabilities.             \\
5   & When 			the CU receives data from Satellites, it shall store the data as 			backup. \\
6   & Sensor 			nodes shall be able to operate with different types of sensors.           \\ \bottomrule
\end{tabularx}%
\vspace{-10px}
\end{table}

For the Data Infrastructure (cf. Table~\ref{tab:2-key-requirements-data-infrastructure}), we first specify that the
system must provide an API, such that new data can be uploaded and
integrated into a Frontend system, allowing users to view and download
the collected data. Additionally, we define the need for different views
for different kinds of users, i.e., easy to grasp summaries for regular
users, and sophisticated time-series graphs for experts. Two further
requirements specify that all data entering the system must undergo
validity checks, and that parameters of these checks must be
configurable on a per-sensor basis by an administrator. Moreover, the
raw measurement values must always be preserved. This allows for
additional calculations or validity checks on the data by experts.
Lastly, the entire system shall be configurable by an administrator.
This means that an administrator can add, remove, or modify (1) sensor
nodes including their connected sensors and their validity check
parameters, (2) FORTE-WSN deployments, and (3) users that have access to
the collected data.

\begin{table}[]
\centering
\caption{Key Requirements for the Data Infrastructure.}
\label{tab:2-key-requirements-data-infrastructure}
\begin{tabularx}{\columnwidth}{lX}
\toprule
\textbf{Nr.} & \textbf{Data Infrastructure}                                                                                                    \\ \midrule
1   & The 			system shall provide an API for interfacing with different 			applications.                                        \\
2   & Authorized 			users shall be able to view and retrieve collected data.                                                    \\
3   & When 			a data point is added to the system, the system shall validate the 			data based on configurable validity checks. \\
4   & The 			integrity of raw data shall be preserved.                                                                          \\
5   & Administrators 			shall be able to configure the system and its deployments.                                              \\ \bottomrule
\end{tabularx}%
\vspace{-10px}
\end{table}
\section{The FORTE System}
\label{sec:forte-infrastructure} 
\begin{figure*}
    \centering
    \includegraphics[width=1\linewidth]{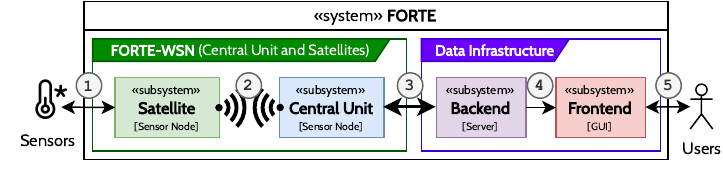}
    \vspace{-1.5em}
    \caption{High-Level Architecture Overview of the FORTE System.}
    \label{fig:1-forte-architecture-overview}
\end{figure*}

The FORTE system contains two main components: (1) the FORTE-WSN
component capable of monitoring a forest environment using its
Satellites and the CU, and (2) the Data Infrastructure that receives,
processes, stores, and presents the data collected by FORTE-WSN (cf.
Figure~\ref{fig:1-forte-architecture-overview}).

\hypertarget{architecture}{%
\subsection{Architecture}\label{architecture}}

\textbf{FORTE-WSN.} The key feature of the FORTE-WSN architecture is the
decision to split the in-forest measurement stations into a Central Unit
(CU), providing long-range communication and sensing capabilities, and
several spatially independent Satellites, that take individual
measurements (Figure~\ref{fig:1-forte-architecture-overview}, \Circled{1}) which are then wirelessly transferred to the
Central Unit (Figure~\ref{fig:1-forte-architecture-overview}, \Circled{2}). Together, the CU and the Satellites thus form
a WSN. Each sensor node (i.e., Satellite or logger at the CU) collects
data using one or more connected sensors. The sink node (i.e., CU)
receives data from the sensor nodes and transmits it to an off-site
location using mobile networking technologies~\cite{matinOverviewWirelessSensor2012}. Both
Satellites and the CU use ESP32 microcontrollers as their processing
units, which allows us to use ESP-NOW\footnote{\url{https://www.espressif.com/en/solutions/low-power-solutions/esp-now},
  accessed on Oct. 30, 2024.} for intra-network communication in the
field. Communication from the CU to the Backend (Figure~\ref{fig:1-forte-architecture-overview}, \Circled{3}) is done
using LTE-M\footnote{Long-Term Evolution Machine Type Communication
  (LTE-M, also known as LTE Cat M1).}. Although reducing data
throughput, {LTE-M} allows for power-efficient data transmission, making
it a good fit for IoT devices~\cite{akpakwuSurvey5GNetworks2018}.

\textbf{Data Infrastructure.} The Data Infrastructure consists of two
subsystems: (1) a Backend server responsible for receiving, processing,
and storing incoming data from the CU of the FORTE-WSN component, and a
(2) Frontend (UI) for users of the system. The Backend includes built-in
data-quality analysis routines that ensure the reliability of incoming
data. These routines include, among others, outlier detection, missing
value detection, and frozen value detection. Users can configure these
validation routines via the Frontend. Upon receiving data, the Backend
automatically applies the configured validation routines and augments
each data point with quality flags that indicate if the received data is
reliable, while preserving the original raw values. Once stored, the
data is accessible via an API. Our Frontend application leverages the
API (Figure~\ref{fig:1-forte-architecture-overview}, \Circled{4}) to provide pre-designed visualizations tailored to
user-specific groups and a download function (Figure~\ref{fig:1-forte-architecture-overview}, \Circled{5}). Besides
retrieving data, the Backend also provides time synchronization services
to the CU. The FORTE system also allows substitution of the Backend with
alternative platforms that allow integration of sensor data via HTTP.

\subsection{Demonstrator Setup}\label{sec:demonstrator-setup}

\begin{figure*}[th!]
    \centering
    \includegraphics[width=1\linewidth]{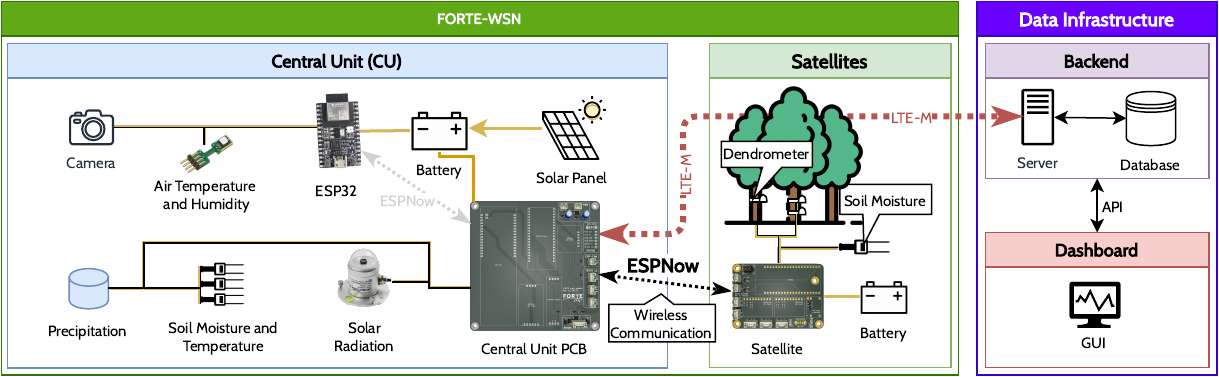}
    \vspace{-1.5em}
    \caption{High-level Overview of the Demonstrator System.}
    \label{fig:2-demonstrator-overview}
\end{figure*}

\begin{figure*}
    \centering
    \includegraphics[width=1\linewidth]{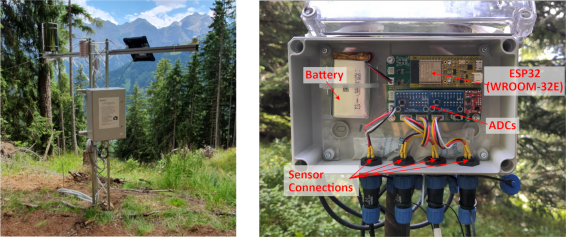}
    \vspace{-1.5em}
    \caption{Image of a Central Unit (left) and a Satellite (right) of the Demonstrator.}
    \label{fig:3-central-unit-demonstrator-picture}
\end{figure*}

Based on our proposed architecture, we built a demonstrator system to
evaluate if it is capable of our goal of performing measurements over
several weeks in a forest microclimate (cf. Figure~\ref{fig:2-demonstrator-overview}). A single
demonstrator consists of two Satellites and a CU. The CU contains the
necessary hardware for communication within the field and the Data
Infrastructure. Additionally, it contains two of the loggers also
utilized in our Satellites to read sensors located close to the CU
itself. Each Satellite is connected to three dendrometers and one
soil-moisture sensor, whereas the rest of the sensors (listed in Table~\ref{tab:3-component-prices}) are connected to loggers at the CU. Figure~\ref{fig:3-central-unit-demonstrator-picture} depicts a CU (left) and
a Satellite (right). The specific electronic components for the loggers
(both CU and Satellites) and other materials used in the demonstrator
are listed in Table~\ref{tab:3-component-prices}.

We deployed three demonstrators at Neustift im Stubaital, Tyrol, at
three different altitudes of 1000 m, 1500 m, and 2000 m respectively,
for a duration of about 3 months. The topmost deployment was placed in
the ``Long-Term Ecological Research (LTER) Master Site Stubai Valley''~\cite{oberleitner2022Amplifying}, which allowed us to compare the data retrieved from our
deployment to trusted values from research-grade sensors of Ecology.

\section{Evaluation} 
\label{sec:evaluation}

We evaluate our FORTE system in two ways: First, we demonstrate cost and
energy efficiency of our solution (cf.~\citesec{efficiency}). Second, we assess
the accuracy of our solution by comparing the data collected at the LTER
master site against a baseline dataset from Ecology and analyze observed
packet loss (cf.~\citesec{comparison}).

\subsection{Cost and Energy
Efficiency}\label{sec:efficiency}

One key requirement is the low component cost of our FORTE-WSN system.
The majority of costs associated with our measurement system come from
individual sensors. Per deployment (one CU, two Satellites), we use a
total of 14 environmental sensors, resulting in a cost of about €2700
(cf. Table~\ref{tab:3-component-prices} for the exact choice of sensors). Using fewer or cheaper
sensors can reduce this price: other I\textsuperscript{2}C, RS-485,
SDI-12, 1-wire, or analog sensors can be used instead (the Satellites
support sensors requiring up to 5V, whereas 5V and 12V are available at
the CU). However, the most significant cost benefits compared to
commercially available systems come from the prices of our CU and
Satellite loggers, which constitute a small fraction of the usual market
prices (cf. Table~\ref{tab:3-component-prices}).

Another key quality requirement of the system is energy efficiency.
In-lab experiments have shown that power consumption of the Satellites
is highest during wireless transmission of data. To optimize the number
of required transmissions per measurement cycle, we implemented a data
structure that merges up to 14 measurements including their metadata
into a single transmission. Additionally, we leverage the deep-sleep
capabilities of the ESP32 platform. This significantly reduces power
consumption, enabling several months of continuous operation on a
2000mAh LiPo battery. The CU requires more energy as it is always on,
ready to receive messages from nearby Satellites. It also contains an
LTE-M modem: we measured a mean current of \textasciitilde175mA over a
32-second data transmission period, dropping to \textasciitilde120mA
when idling for 30 minutes after the transmission has finished.
Depending on the amount of data to be sent, the transmission period may
increase. To provide continuous power at the CU, we use a 17Ah lead
battery installed in the loggerbox, allowing operation for about a week
without energy harvesting.

\begin{table}[hb]
\centering
\renewcommand{\arraystretch}{1.25}
\caption{List of demonstrator components and their approximate prices.}
\label{tab:3-component-prices}
\begin{tabular}{L{1.7cm}r L{4.9cm}} %
\toprule
Category                       & Price      & Components                                                                                                                                                                                     \\ \midrule
CU 			– logger                 & €100      

&\begin{minipage}{1.0\linewidth}			custom-designed PCB, containing an  LTE-M module, ESP32, etc.
\end{minipage}                                                                                  \\
CU 			– other materials        & €540       & Solar 			panel, solar regulator, 17Ah lead acid battery, enclosure, trusses 			and fixtures, etc.                                                                                              \\
Satellite 			– logger          & 2 			x €25 & PCB: 			ESP32, ADCs                                                                                                                                                                            \\
Satellite 			– other materials & €115       & LiPo 			battery, enclosure, connectors, fixtures                                                                                                                                               \\
Sensors                        & €2700     
&\begin{minipage}{1.0\linewidth}
1x 			air temperature and humidity,\\ 3x soil temperature and moisture, 2x 			soil moisture,\\ 1x solar radiation,\\ 1x precipitation,\\  6x 			dendrometer
\end{minipage}
\\ \bottomrule
\end{tabular}%
\vspace{-10px}
\end{table}

\subsection{Data Comparison and Package
Loss}
\label{sec:comparison}

To assess the accuracy of our demonstrator, we collected and compared
data from our FORTE-WSN system with an operational, scientific grade
measurement system running at the LTER master site Stubai Valley~\cite{oberleitner2022Amplifying}. This site offered the opportunity to directly compare
different types of dendrometers (band and point) measuring radial growth
of the trees. For the work case of our demonstrator, the dendrometers
were the key instrument installed. Therefore, only these measurements
are shown in the data comparison.

\begin{figure}
    \centering
    \includegraphics[width=\columnwidth]{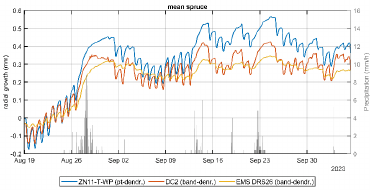}
    \includegraphics[width=\columnwidth]{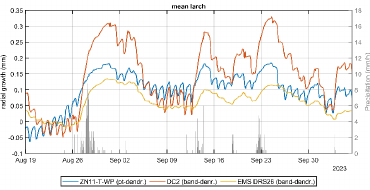}
    \vspace{-1.5em}
    \caption{Mean radial growth of all the examined spruces (top) and
larches (bottom) in the phase of parallel measurements using point and
band dendrometers, as well precipitation time series. Shown are averaged
values per tree species and measuring device.}
    \label{fig:4-data-comparison}
\end{figure}

\textbf{Data Comparison.} For the intercomparison, 3 spruces and 3
larches were sampled, which were already equipped with point (ZN11-T-WP,
Natkon) and band dendrometers (DC2, Ecomatik GmbH). We then added our
FORTE-WSN band dendrometers (EMS DR26, EMS Brno) to these same
individuals. This allowed a direct comparison of the band dendrometers
used in FORTE-WSN with measurements using the same principle and an
alternative method (point dendrometer) on the same individuals (cf.
Figure~\ref{fig:4-data-comparison}). The comparison period is in the stagnation phase of tree
growth (i.e., the main yearly growing phase is over). This leads to a
weak signal, which is difficult to resolve by measurements. Still, we
were able to capture an initial growth signal, especially in the spruces
(19.-26.08.2023), as well as phases with swelling tree trunks due to
humid weather conditions (e.g., 26.08.2023--02.09.2023) and subsequent
drying phases. In general, the band dendrometers (EMS DR26) show good
agreement with the point and band dendrometers already installed on the
same trees.

\textbf{Packet Loss.} Here, we analyze how many packets that were sent
from each Satellite, and relayed over the CU, arrived at the Backend
server. We consider the raw readings collected by our three
demonstrators between the 30 days of September 3rd, 2023, and October
3rd, 2023. We set up our Satellites to provide 2 readings per sensor per
hour, thus expecting one measurement per sensor every 30 minutes (1800
seconds). Due to the low accuracy and high-temperature dependency of the
ESP32 sleep timer, we get a mean and standard deviation of 1790$\pm$9
seconds per measurement. Note that this clock drift is of no consequence
for our application and could be reduced using synchronization
techniques. During the evaluation period, the Satellites reached a
packet loss of under 1\%, and no Satellite missed more than 3
transmissions in a row. However, the package loss is highly
condition-dependent (terrain, Satellite-CU distance, vegetation
obscuring the line of sight, etc.), and could increase under less ideal
circumstances.

\subsection{Discussion}\label{discussion}

Our evaluation shows that, in the use case of our forest monitoring
demonstrator, the FORTE system achieved data accuracy and reliability
comparable to industry-standard measurements. Given the flexibility of
the FORTE system, this setup could be adapted for a variety of
environmental monitoring applications using different sensors. The
innovative configuration of a CU with multiple Satellites provides new
opportunities for versatile and scalable environmental monitoring.
Additionally, the 3-month-long field test allowed us to monitor the
system in a real-world environment. This gave us additional insights for
further improvement. Several problems were found and solved during the
real-world test, from hardware bugs on the PCBs to problems with data
transmission. Notably, no data was lost during connection issues, which
demonstrates the efficacy of our backup system.

After the initial 3-month field test, we redeployed a station on
November 21st, 2023, at a test site of the Tyrolean Forest Service near
Rechenhof in Innsbruck. This station gave us a better understanding of
how the station operates during winter conditions and over longer
periods of time. With the station, we were able to verify the energy
efficiency requirements of our Satellites. Both Satellites reported a
battery status of around 10\% after 11 months of operation. However, we
have also noticed that packet loss increases in Spring, likely due to
increased vegetation density, which is why we aim to use alternative
network communication methods such as low-frequency LoRa in future work.
\section{Conclusion / Future Work} \label{sec:conclusion}

In this paper, we presented the key requirements, architecture, and
prototype implementation of the FORTE system for environmental
monitoring. We showed that our solution is flexible and cost-effective,
with individual Satellite loggers costing around €25 and a CU logger
around €100. We deployed three separate stations over a duration of 3
months to evaluate the effectiveness of our system. Additionally, we
redeployed a station to gain better insights on how the system performs
during the months of Winter and Spring. Future work will focus on
further reducing power consumption and exploring alternative network
communication means within the forest.

 \section*{Acknowledgment}
We thank Anurag Vats for implementing the Backend, as well as Bilal
Hassan for his work on the Frontend. Further, we thank Moritz Perschke
for helping with the initial prototyping phase of the sensor network. We
thank Michael Bahn for making the dendrometer data from the LTER site
Stubai Valley available for comparison against our system. This study
was part of the project: ``FORTE: Offene, skalierbare Daten für
evidenzbasierte Entscheidungen im Wald der Zukunft''. The project was
funded by the Federal Ministry of Agriculture, Forestry, Regions and
Water Management of the Republic of Austria (BML) through the DaFNE
program (grant no. 101699).

\normalsize

\bibliography{FORTE} 
\balance

\end{document}